\documentclass[twocolumn,prb,showpacs]{revtex4}%
\usepackage{graphicx}%
\usepackage{amsmath}%
\usepackage{amsfonts}%
\usepackage{amssymb}
\usepackage{bm}

\def\q{{ {\bm q} }}

\def\w{{\omega}}
\def\a{{\alpha}}

\def\g{{\gamma}}

\allowdisplaybreaks[4]

\begin{document}
\title{
Emergence of Fully-Gapped $s_{++}$-wave and Nodal $d$-wave States Mediated by \\
Orbital- and Spin-Fluctuations in Ten-Orbital Model for KFe$_2$Se$_2$
}
\author{
Tetsuro \textsc{Saito}$^{1}$, Seiichiro \textsc{Onari}$^{2}$, 
and Hiroshi \textsc{Kontani}$^{1}$}
\date{\today }

\begin{abstract}
We study the superconducting state in 
recently discovered high-$T_{\rm c}$ superconductor K$_x$Fe$_2$Se$_2$
based on the ten-orbital Hubbard-Holstein model without hole-pockets.
When the Coulomb interaction is large,
spin-fluctuation mediated $d$-wave state appears
due to the nesting between electron-pockets.
Interestingly, the symmetry of the body-centered tetragonal structure
in K$_x$Fe$_2$Se$_2$ requires the existence of nodes in the $d$-wave gap,
although fully-gapped $d$-wave state is realized
in the case of simple tetragonal structure.
In the presence of moderate electron-phonon interaction 
due to Fe-ion optical modes, on the other hand,
orbital fluctuations give rise to the 
fully-gapped $s_{++}$-wave state without sign reversal.
Therefore, both superconducting states are distinguishable 
by careful measurements of the gap structure
or the impurity effect on $T_{\rm c}$.
\end{abstract} 

\address{
$^1$ Department of Physics, Nagoya University and JST, TRIP, 
Furo-cho, Nagoya 464-8602, Japan. 
\\
$^2$ Department of Applied Physics, Nagoya University and JST, TRIP, 
Furo-cho, Nagoya 464-8602, Japan. 
}
 

\sloppy

\maketitle


The pairing mechanism of high-$T_{\rm c}$ iron-based superconductors 
has been significant open problem.
The main characters of FeAs compounds would be
(i) the nesting between electron-pockets (e-pockets)
and hole-pockets (h-pockets), and 
(ii) the existence of orbital degree of freedom.
By focusing on the intra-orbital nesting, 
fully-gapped sign-reversing $s$-wave state 
($s_\pm$-wave state) had been predicted based on the 
spin fluctuation theories
 \cite{Mazin,Kuroki}.
On the other hand, 
existence of moderate electron-phonon ($e$-ph) interactions 
due to Fe-ion optical phonons and the inter-orbital nesting
can produce large orbital fluctuations 
\cite{Kontani}.
Then, orbital-fluctuation-mediated $s$-wave state without sign reversal 
($s_{++}$-wave state) had been predicted by using the 
random-phase-approximation (RPA) \cite{Kontani,Saito}
or the fluctuation-exchange (FLEX) approximation \cite{Onari}.
According to the analysis in Refs. \onlinecite{Onari-impurity,Onari-resonance}
the $s_{++}$-wave state is consistent with the robustness of $T_{\rm c}$
against randomness \cite{Sato-imp,Nakajima} 
as well as the ``resonance-like'' hump structure 
in the neutron inelastic scattering \cite{neutron}.
Non-Fermi liquid transport phenomena in $\rho$ 
\cite{Matsuda}
can be explained by the development of orbital fluctuations
\cite{Onari}.

Recently, iron-selenium 122-structure compound A$_x$Fe$_2$Se$_2$ 
(A= alkaline metals) with $T_{\rm c}\sim30$ K was discovered
 \cite{first}.
This heavily electron-doped superconductor
has been attracting great attention since
both the band calculations \cite{LDA1,LDA2}
and angle-resolved-photoemission-spectrum (ARPES) measurements 
\cite{ARPES1,ARPES2,ARPES3} indicate the absence of h-pockets.
NMR measurements reports the weakness of spin fluctuations
 \cite{NMR},
and both ARPES \cite{ARPES1,ARPES2,ARPES3} and
specific heat measurements \cite{HHWen} indicate the isotropic SC gap.
Thus, study of A$_x$Fe$_2$Se$_2$ will give us important 
information to reveal the pairing mechanism of iron pnictides.

The unit-cell of iron-based superconductors contains two Fe atoms.
However, except for 122-systems, 
one can construct a simple ``single-Fe model'' from the 
original ``two-Fe model'' by applying the
gauge transformation on $d$-orbitals \cite{Miyake}.
By this procedure, the original Brillouin zone (BZ) is 
enlarged to the ``unfolded BZ''.
Based on the single-Fe model, 
spin-fluctuation-mediated $d$-wave state ($B_{1g}$ representation)
``without nodes'' had been proposed
 \cite{DHLee,Graser,Bala},
by focusing on the nesting between e-pockets.
However, we cannot construct a ``single-Fe model''
for 122 systems since finite hybridization between 
e-pockets prevents the unfolding procedure 
\cite{Miyake}.
Therefore, theoretical study based on the original 
two-Fe model is highly desired to conclude the gap structure.

In this paper, we study the ten-orbital (two Fe atoms) 
Hubbard-Holstein (HH) model for KFe$_2$Se$_2$ using the RPA.
When the Coulomb interaction is large,
we obtain the $d$-wave SC state due to the spin fluctuations,
as predicted by the recent theoretical studies
in the single-Fe Hubbard models \cite{DHLee,Graser,Bala}.
However, the gap function on the Fermi surfaces (FSs) 
inevitably has ``nodal structure'' in the two-Fe model,
due to the symmetry requirement of the 
body-centered tetragonal lattice.
On the other hand, 
orbital-fluctuation-mediated $s_{++}$-wave state is realized 
by small $e$-ph coupling;
the dimensionless 
coupling constant $\lambda=gN(0)$ is just $\sim0.2$.
Since the nodal SC state is fragile against randomness,
study of impurity effect will be useful to distinguish these SC states.

\begin{figure}[!htb]
\includegraphics[width=0.99\linewidth]{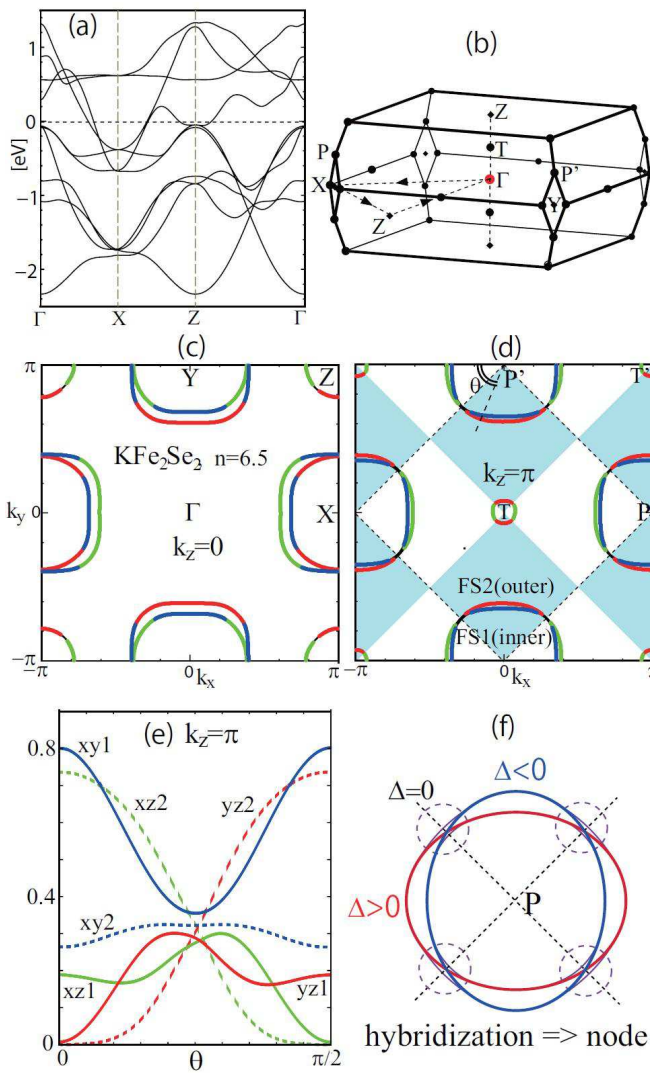}
\caption{(color online) 
(a) Dispersion of the present ten-orbital model for KFe$_2$Se$_2$.
$\Gamma$, $X$, and $Z$ points are on the $k_z=0$ plane.
(b) Primitive BZ for body-centered tetragonal lattice.
(c)(d) FSs on the $k_z=0$ plane and $k_z=\pi$ plane.
The green (light gray), red (gray), and blue (dark gray) lines
correspond to $xz$, $yz$, and $xy$ orbitals, respectively.
The diamond-shaped shadows in (d) indicates the sign of 
basis function for $B_{1g}$ representation.
(e) Weight of each orbital on the inter FS (FS1) and 
the outer FS (FS2) as function of $\theta$; $\theta$ is shown in (d).
(f) Hybridization between two e-pockets
in 122 systems should create the nodal $d$-wave gap.
}
\label{fig:fig1}
\end{figure}

We perform the local-density-approximation (LDA) band calculation 
for KFe$_2$Se$_2$ using Wien2k code based on the experimental 
crystal structure \cite{first}.
Next, we derive the ten-orbital tight-binding model that 
reproduces the LDA band structure and its orbital character
using Wannier90 code and Wien2Wannier interface \cite{Arita}.
The dispersion of the model and the primitive BZ are
shown in Figs. \ref{fig:fig1} (a) and (b).
Based on a similar ten-orbital model, Suzuki {\it et al.} studied 
the $s_\pm$-wave gap structure for BaFe$_2$As$_2$ \cite{Suzuki}.

In Fig. \ref{fig:fig1}, we show the FSs of KFe$_2$Se$_2$ 
for (c) $k_z=0$ and (d) $k_z=\pi$ planes
when the electron number per Fe-ion is $n=6.5$:
On each plane, there are four large and heavy e-pockets around X and Y points,
and one small and light e-pockets around Z point.
For $n=6.5$, the energy of the h-band at $\Gamma$ point
from the Fermi level, $E_h$, is about $-0.07$ eV.
Since the obtained FS topology and the value of $E_h$ are consistent with 
recent reports by ARPES measurements \cite{ARPES1,ARPES2,ARPES3},
we study the case $n=6.5$ hereafter.
In the present BZ in (b),
$\Gamma$ and Z points and X and Y points in (c) are not equivalent,
and $k_z=2\pi$ plane is given by shifting (c) by $(\pi,\pi)$.
As for (d), T and T' points and P and P' points are equivalent,
meaning that the reciprocal wave vector 
on the $k_z=\pi$ plane is $(\pi,\pi)$ and $(\pi,-\pi)$.
The diamond-shaped shadows in the $k_z=\pi$ plane 
indicates the sign of basis function
for $B_{1g}$ ($x^2-y^2$-type) representation, which has
nodes on the P-P' line on both FS1 (inner FS) and FS2 (outer FS).

To confirm the existence of nodes, we verify that 
FS1 and FS2 in KFe$_2$Se$_2$ are largely hybridized.
In fact, the weights of $d$-orbitals on FS1,2 given in Fig. \ref{fig:fig1} (e)
are smooth functions of $\theta$, which is the strong evidence for
the hybridization in wide momentum space.
This hybridization disappears when inter-layer hoppings are neglected: 
Then, both $xy({\rm FS1})$ and $xy({\rm FS2})$
show cusps at $\theta=\pi/4$, and $xz({\rm FS2})$ 
suddenly drops to almost zero for $\theta\ge\pi/4$.
In (f), we explain the origin of nodal gap based on the fully-gapped 
$d$-wave solution in the single-Fe model \cite{DHLee,Graser,Bala}:
By introducing inter-layer hoppings,
two elliptical e-pockets with positive and negative $\Delta$
in the unfolded BZ are hybridized to form FS1 and 2
with four-fold symmetry.
As a result, nodal lines inevitably emerge on FS1 and 2,
at least near the $|k_z|=\pi$ plane.

Here, we study the ten-orbital HH model using the RPA.
As for the Coulomb interaction, we consider the intra-orbital term $U$, 
the inter-orbital term $U'$, Hund's coupling or pair hopping $J$,
and assume the relation $U=U'+2J$ and $J=U/6$.
In addition, we consider the $e$-ph interaction 
due to Fe-ion optical phonons;
the phonon-mediated e-e interaction ($-g$) and its matrix elements
are presented in Ref. \onlinecite{Saito}.
Hereafter, we perform the RPA on the two-dimensional planes
for $k_z=0$, $\pi/2$, and $\pi$.

\begin{figure}[!htb]
\includegraphics[width=0.99\linewidth]{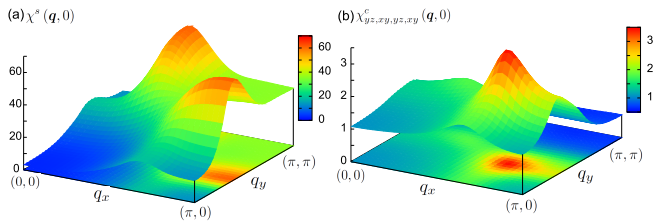}
\caption{(color online) 
(a) $\chi^s(\q,0)$ for $U=1.1$ eV and $g=0$, and
(b) $\chi^c_{yz,xy;yz,xy}(\q,0)$ for $U=0$ and $g=0.22$ eV.
}
\label{fig:fig2}
\end{figure}

For $n=6.5$ and $k_z=0$, the critical value of $g$ 
for the orbital-density-wave (ODW) is $g_{\rm c}=0.23$ eV for $U=0$,
and the critical value of $U$ for the spin-density-wave (SDW) 
is $U=1.18$ eV for $g=0$. 
These values change only $\sim2\%$ for different $k_z$.
The obtained $U$-$g$ phase diagram is very similar to 
Fig. 2 in Ref. \onlinecite{Saito}, irrespective of the absence of
h-pockets in KFe$_2$Se$_2$.
The reason would be (i) the density-of-states (DOS)
in KFe$_2$Se$_2$ is about 1eV$^{-1}$ per Fe, which is 
comparable with other iron pnictides, and 
(ii) the nesting between e-pockets is rather strong
because of their square-like shape.
Figure \ref{fig:fig2} (a) shows the total spin susuceptibility 
$\chi^s(\q,0)$ at $U=1.1$ eV and $g=0$ for $k_z=0$ plane.
$\chi^s$ is given by the intra-orbital nesting,
and its peak position is $\q\approx(\pi,0.4\pi)$, consistenlty 
with previous studies \cite{DHLee,Graser,Bala}.
The obtained incommensulate spin correlation
is the origin of the $d$-wave SC gap.
Figure \ref{fig:fig2} (b) shows the off-diagonal 
orbital susceptibility $\chi_{yz,xy;yz,xy}^c(\q,0)$ for the $k_z=0$ plane
at $U=0$ and $g=0.22$ eV;
its definition is given in Refs. \onlinecite{Kontani,Saito,Onari}.
It is derived from the inter-orbital nesting
between $xz$ and $xy$, and its peak position is $\q\approx(0.7\pi,0.4\pi)$.
Note that the peak position of $\chi_{xz,xy;xz,xy}^c$ is $\q\approx(0.4\pi,0.7\pi)$.
The obtained strong spin- and orbital-correlations
are the origin of the $d$-wave and $s_{++}$-wave SC states.

\begin{figure}[!htb]
\includegraphics[width=0.99\linewidth]{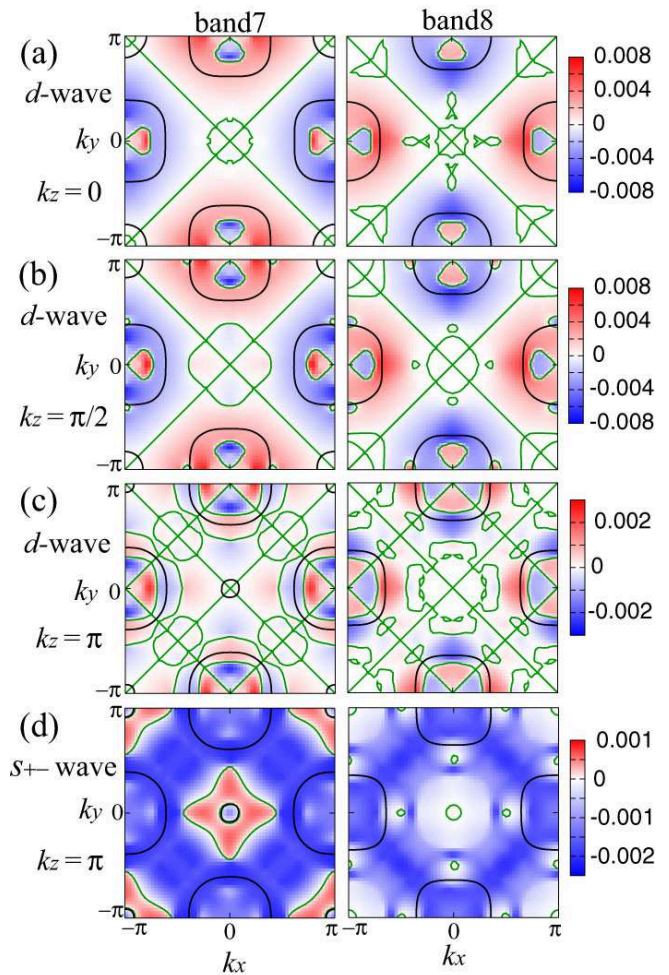}
\caption{(color online) 
SC gap functions for outer FS (FS2) on band 7 and inner FS (FS1) on band 8:
$d$-wave gap functions on the (a) $k_z=0$, (b) $k_z=\pi/2$, and 
(c) $k_z=\pi$ planes.
Black and green (gray) lines represent the FSs and gap nodes.
(d) $s_\pm$-wave gap function on the $k_z=\pi$ plane.
}
\label{fig:fig3}
\end{figure}

In the following, we solve the linearized gap equation to 
obtain the gap function,
by applying the Lanczos algorithm to achieve reliable results.
In the actual calculation results shown below, we take 
$64\times64$ ${\bm k}$-point meshes and 512 Matsubara frequencies.
First, we study the spin-fluctuation-mediated SC state
for $U\lesssim U_{\rm c}$ by putting $g=0$.
Figures \ref{fig:fig3} (a)-(c) show the gap functions
of the $d$-wave solution at $T=0.03$ eV
for $k_z=0$, $\pi/2$, and $\pi$, respectively.
In case of $U=1.1$ eV, the eigenvalue $\lambda_{\rm E}$ is
$0.61$ for (a), $0.63$ for (b), and $0.62$ for (c);
the relation $\lambda_{\rm E}\ge1$ corresponds to the SC state.
They are relatively small since the SC condensation energy
becomes small when the SC gap has complicated nodal line structure.
On the (c) $k_z=\pi$ plane, the nodal lines are along $\theta=\pi/4$
and $3\pi/4$ directions, consistently with the 
basis of $B_{1g}$ representation in Fig. \ref{fig:fig1} (d).
These nodes move to near the BZ boundary, $\theta=0$ and $\pi$,
on the (b) $k_z=\pi/2$ plane,
and they deviate from the FSs on the (a) $k_z=0$ plane.
As results, the nodal gap 
appears for $\pi/2<|k_z|<3\pi/2$ in the whole BZ $|k_z|\le 2\pi$.

We also obtain the $s_\pm$-wave state, with the sign reversal
of the SC gap between e-pockets and the 
``hidden h-pockets below the Fermi level'' given by the valence bands 5,6.
The obtained solution is shown in Fig. \ref{fig:fig3} (d) for $k_z=\pi$.
Interestingly, the obtained eigenvalue is $\lambda_{\rm E}=0.99$
for $U=1.1$ eV,
which is larger than $\lambda_{\rm E}$ for $d$-wave state in 
Fig. \ref{fig:fig3} (a)-(c).
Such large $\lambda_{\rm E}$ originates from the scattering of Cooper pairs
between e-pockets and the ``hidden h-pockets'',
which was discussed as the ``valence-band Suhl-Kondo (VBSK) effect''
in the study of Na$_x$CoO$_2$ in Ref. \onlinecite{Yada}.

\begin{figure}[!htb]
\includegraphics[width=0.99\linewidth]{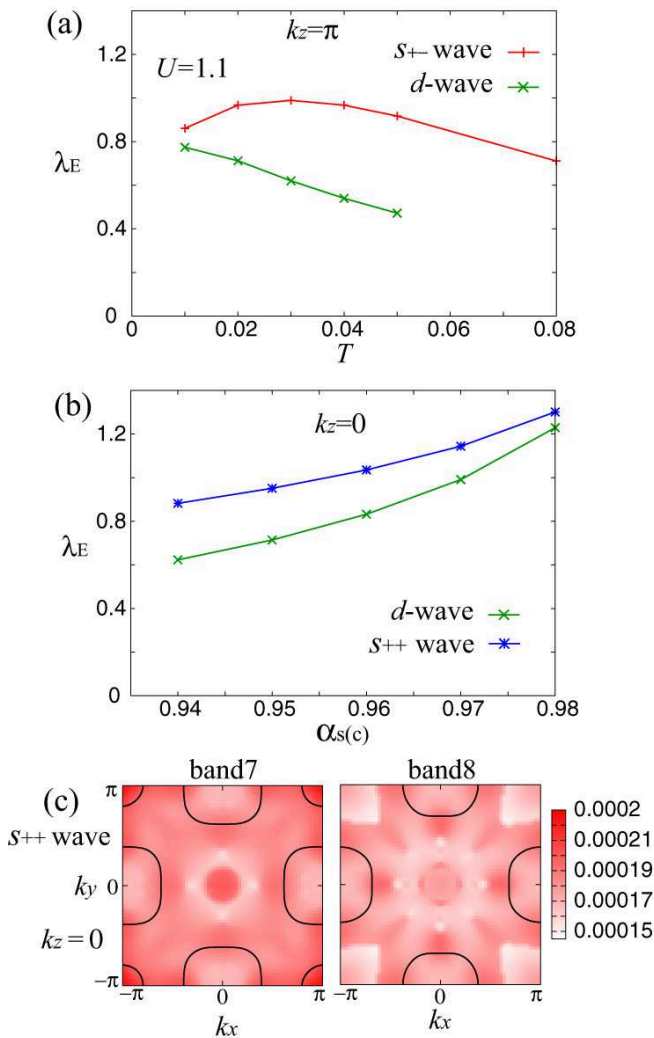}
\caption{(color online) 
(a) $T$-dependence of $\lambda_{\rm E}$ for $d$- and $s_\pm$-wave states.
$\lambda_{\rm E}$ at $T=0.01$ eV is underestimated because of 
the shortage of ${\bm k}$- and Matsubara-meshes.
(b) $\a_{s}$- ($\a_c$-) dependence of $\lambda_{\rm E}$
for $d$-wave ($s_{++}$-wave) state at $T=0.03$ eV.
(c) SC gap functions for $s_{++}$-wave state.
}
\label{fig:fig4}
\end{figure}

Here, we analyze the $T$-dependence of $\lambda_{\rm E}$
based on a simple two-band model with inter-band repulsion:
The set of gap equations is given by \cite{Yada}
$\lambda_{\rm E}\Delta_h = -VN_e L_e\Delta_e$ and
$\lambda_{\rm E}\Delta_e = -VN_h L_h\Delta_h$,
where $V>0$ is the repulsive interaction between e- and h-pockets,
and $N_{e,h}$ is the DOS near the Fermi level.
When (i) the top of the h-pocket is well above the Fermi level,
$L_e=L_h=\ln(1.13\w_c/T)$, where $\w_c$ is the cutoff energy.
Thus, the eigenvalue is given as
$\lambda_{\rm E}=V\sqrt{N_eN_h}\ln(1.13\w_c/T)\propto -\ln T$,
similar to single-band BCS superconductors.
On the other hand, when (ii) h-pocket is slightly below the Fermi level, 
$L_h=(1/2)\ln(\w_c/|E_h|)$,
where $E_h<0$ is the energy of the top of h-band \cite{Yada}.
Thus, the eigenvalue is given as
$\lambda_{\rm E}=V\sqrt{N_eN_hL_h}\sqrt{\ln(1.13\w_c/T)}\propto \sqrt{-\ln T}$.
Therefore, in case (ii), the $T$-dependence of $\lambda_{\rm E}$ 
is much moderate.
In fact, as shown in Fig. \ref{fig:fig4} (a),
$\lambda_{\rm E}$ for $d$-wave state increases monotonically
with decreasing $T$, while $\lambda_{\rm E}$ for $s_\pm$-wave state
saturates at low temperatures.
This result suggests that the $d$-wave state overcomes the 
$s_\pm$-wave state at $T_{\rm c}\sim30$K in K$_x$Fe$_2$Se$_2$.
Although $T_{\rm c}$ in the $s_\pm$-wave state is $\sim 0.06$ eV
in Fig. \ref{fig:fig4} (a), it is greatly reduced by the self-energy 
correction that is absent in the RPA \cite{Onari}.

We discuss the VBSK effect for $s_\pm$ wave state in more detail:
According to inelastic neutron scattering measurement
of Ba(Fe,Co)$_2$As$_2$ \cite{neutron},
the characteristic spin-fluctuation energy is 
$\w_{\rm sf}\sim 100$K just above $T_{\rm c}\sim30$K.
If we assume a similar $\w_{\rm sf}$ in KFe$_2$Se$_2$
since $T_{\rm c}$ is close,
we obtain the relation $\w_c\sim\w_{\rm sf}\ll |E_h|$ in KFe$_2$Se$_2$.
Since $L_h$ is a monotonic decrease function of 
$|E_h|/\w_c$ and $L_h<1$ for $-E_h/\w_c>0.15$,
we consider that $d$-wave state overcomes the 
$s_\pm$-wave state in KFe$_2$Se$_2$, as far as the 
spin-fluctution mediated superconductivity is considered.
Although high-$T_{\rm c}$ $s_\pm$-wave state mighe be realized 
for $|E_h|/\w_c<0.1$, then the realized $T_{\rm c}$ will be 
very sensitive to $E_h$ or the filling $n$ \cite{Yada}.


Now, we study the $s_{++}$-wave state due to orbital fluctuations
on the $k_z=0$ plane with $n=6.5$.
In Fig. \ref{fig:fig4} (b), we show the $\a_c$-dependence of 
$\lambda_{\rm E}$ at $T=0.03$ for the $s_{++}$-wave state with $U=0$,
and the $\a_s$-dependence of 
$\lambda_{\rm E}$ for the $d$-wave state with $g=0$.
Here, $\a_c$ ($\a_s$) is the charge (spin) Stoner factor 
introduced in Ref. \onlinecite{Kontani};
$\a_c=1$ ($\a_s=1$) corresponds to the ODW (SDW) state. 
In calculating the $s_{++}$-wave state,
we use rather larger phonon energy; $\w_D=0.15$ eV,
considering that the calculating temperature is about ten times 
larger than the real $T_{\rm c}$.
The SC gap functions for $s_{++}$-wave state are rather 
isotropic, as shown in Fig. \ref{fig:fig4} (c).
However, the obtained SC gap becomes
more anisotropic in case of $U>0$ \cite{Onari}.

We stress that the RPA is insufficient for quantitative study of 
$\lambda_{\rm E}$ since the self-energy correction $\Sigma$ is dropped:
In Ref. \onlinecite{Onari}, we have studied the present model
based on ther FLEX approximation,
and found that the critical region with $\a_c\gtrsim0.95$ is enlarged 
by the inelastic staccering $\g={\rm Im}\Sigma$. 
Also, the $\gamma$-induced suppression in $\lambda_{\rm E}$
for $d$- or $s_\pm$-wave states
is more prominent than that for $s_{++}$-wave state, 
since $\gamma$ due to spin fluctuations is 
larger than that due to orbtial fluctuaiotns \cite{Onari}.

Recently, we found the paper by Mazin \cite{Mazin2},
in which the $k_z$ dependence of the nodal $d$-wave gap in Fig. 4
corresponds to Figs. \ref{fig:fig3} (a)-(c) in the present work.

In summary, 
we studied the mechanism of superconductivity in KFe$_2$Se$_2$
based on the ten-orbital HH model without h-pockets.
Similar to iron-pnitide superconductors,
orbital-fluctuation-mediated $s_{++}$-wave state
is realized by small dimensionless $e$-ph 
coupling constant $\lambda=gN(0)\sim0.2$.
We also studied the spin-fluctuation-mediated $d$-wave state, 
and confirmed that nodal lines appear on the large e-pockets,
due to the hybridization between two e-pockets
that is inherent in 122 systems.
Therefore, careful measurements on the SC gap anisotropy
is useful to distinguish these different pairing mechanisms.
Study of impurity effect on $T_{\rm c}$ is also useful
since $d$-wave (and $s_\pm$-wave) state is fragile against impurities.

\acknowledgements
We are grateful to D.J. Scalapino, P. Hirschfeld, A. Chubukov,
Y. Matsuda, and other attandances in the international 
workshop ``Iron-Based Superconductors'' in KITP 2011
for useful and stimulating discussions.
This study has been supported by Grants-in-Aid for Scientific 
Research from MEXT of Japan, and by JST, TRIP.
Numerical calculations were performed using the facilities of the
supercomputer centers in ISSP and Institute for Molecular Science.




\begin{thebibliography}{99}

\bibitem{Mazin}
I. I. Mazin, D. J. Singh, M. D. Johannes, and M. H. Du,
Phys. Rev. Lett. {\bf 101}, 057003 (2008).

\bibitem{Kuroki}
K. Kuroki, S. Onari, R. Arita, H. Usui, Y. Tanaka,
H. Kontani, and H. Aoki,
Phys. Rev. Lett. {\bf 101}, 087004 (2008).

\bibitem{Kontani}
H. Kontani and S. Onari, Phys. Rev. Lett. {\bf 104}, 157001 (2010).

\bibitem{Saito}
T. Saito, S. Onari, and H. Kontani,
Phys. Rev. B {\bf 82}, 144510 (2010) 

\bibitem{Onari}
S. Onari and H. Kontani, arXiv:1009.3882.

\bibitem{Onari-impurity}
S. Onari and H. Kontani, 
Phys. Rev. Lett. {\bf 103} 177001 (2009).

\bibitem{Onari-resonance}
S. Onari, H. Kontani, and M. Sato,
Phys. Rev. B {\bf 81}, 060504(R) (2010) 

\bibitem{Sato-imp}
A. Kawabata, S. C. Lee, T. Moyoshi,
Y. Kobayashi, and M. Sato, J. Phys. Soc. Jpn. {\bf 77} (2008) Suppl. C 103704;
M. Sato, Y. Kobayashi, S. C. Lee,
H. Takahashi, E. Satomi, and Y. Miura, J. Phys. Soc. Jpn. {\bf 79} (2010) 014710;
S. C. Lee, E. Satomi, Y. Kobayashi, and M. Sato, J. Phys. Soc. Jpn. {\bf 79} (2010) 023702.

\bibitem{Nakajima}
Y. Nakajima, T. Taen, Y. Tsuchiya, T. Tamegai, H. Kitamura, and T. Murakami,
arXiv:1009.2848.

\bibitem{neutron}
A. D. Christianson, E. A. Goremychkin, R. Osborn, S. Rosenkranz, M. D. Lumsden, C. D. Malliakas,
I. S. Todorov, H. Claus, D. Y. Chung, M. G. Kanatzidis, R. I. Bewley, and T. Guidi, Nature {\bf 456}, 930 (2008);
Y. Qiu, W. Bao, Y. Zhao, C. Broholm, V. Stanev, Z. Tesanovic, Y. C. Gasparovic, S. Chang, J. Hu,
B. Qian, M. Fang, and Z. Mao, Phys. Rev. Lett. {\bf 103}, 067008 (2009);
D. S. Inosov, J. T. Park, P. Bourges, D. L. Sun, Y. Sidis, A. Schneidewind, K. Hradil, D. Haug, C. T. Lin,
B. Keimer, and V. Hinkov, Nature Physics {\bf 6}, 178 (2010).

\bibitem{Matsuda}
S. Kasahara, T. Shibauchi, K. Hashimoto, K. Ikada, S. Tonegawa, R. Okazaki, H. Shishido, H. Ikeda,
H. Takeya, K. Hirata, T. Terashima, and Y. Matsuda, Phys. Rev. B {\bf 81}, 184519 (2010).

\bibitem{first}
J. Guo, S. Jin, G. Wang, S. Wang, K. Zhu, T. Zhou, M. He, and X. Chen, Phys. Rev. B {\bf 82}, 180520(R) (2010).

\bibitem{LDA1}
I.R. Shein and A.L. Ivanovskii, arXiv:1012.5164.

\bibitem{LDA2}
I. A. Nekrasov, and M. V. Sadovskii, 
Pisma ZhETF, {\bf 93}, 182 (2011).

\bibitem{ARPES1}
Y. Zhang, L. X. Yang, M. Xu, Z. R. Ye, F. Chen, C. He, H. C. Xu, J. Jiang, B. P. Xie, J. J. Ying,
X. F. Wang, X. H. Chen, J. P. Hu, M. Matsunami, S. Kimura, and D. L. Feng,
Nature Materials, doi:10.1038/nmat2981.

\bibitem{ARPES2}
T. Qian, X.-P. Wang, W.-C. Jin, P. Zhang, P. Richard, G. Xu, X. Dai, Z. Fang, J.-G. Guo, X.-L. Chen, and H. Ding,
arXiv:1012.6017.

\bibitem{ARPES3}
L. Zhao, D. Mou, S. Liu, X. Jia, J. He, Y. Peng, L. Yu, X. Liu, G. Liu, S. He, X. Dong, J. Zhang, J. B. He, D. M. Wang,
G. F. Chen, J. G. Guo, X. L. Chen, X. Wang, Q. Peng, Z. Wang, S. Zhang, F. Yang, Z. Xu, C. Chen, X. J. Zhou, arXiv:1102.1057.

\bibitem{NMR}
W. Yu, L. Ma, J. B. He, D. M. Wang, T.-L. Xia, G. F. Chen, arXiv:1101.1017.

\bibitem{HHWen}
B. Zeng, B. Shen, G. Chen, J. He, D. Wang, C. Li, H.-H. Wen, arXiv:1101.5117.

\bibitem{Miyake}
T. Miyake, K. Nakamura, R. Arita, M. Imada,
J. Phys. Soc. Jpn. {\bf 79} 044705 (2010).

\bibitem{DHLee}
F. Wang, F. Yang, M. Gao, Z.-Y. Lu, T. Xiang, D.-H. Lee, arXiv:1101.4390.

\bibitem{Graser}
T. A. Maier, S. Graser, P. J. Hirschfeld, and D. J. Scalapino,
arXiv:1101.4988.

\bibitem{Bala}
T. Das and A. V. Balatsky, arXiv:1101.6056.

\bibitem{Arita}
J. Kunes, R. Arita, P. Wissgott, A. Toschi, H. Ikeda, K. Held, Comp. Phys. Commun. {\bf 181}, 1888 (2010).

\bibitem{Suzuki}
K. Suzuki, H. Usui, and K. Kuroki, J. Phys. Soc. Jpn. 80 013710 (2011).

\bibitem{Yada}
K. Yada and H. Kontani, Phys. Rev. B {\bf 77}, 184521 (2008);
K. Yada and H. Kontani, J. Phys. Soc. Jpn. {\bf 75}, 033705 (2006).

\bibitem{Mazin2}
I. I. Mazin, arXiv:1102.3655.

\end{thebibliography}
\end{document}